\documentclass[aps,prc,amsmath,showpacs,superscriptaddress,nofootinbib]{revtex4}
\usepackage{graphicx,epsfig,epstopdf,longtable}
\usepackage[dvipsnames]{xcolor}
\usepackage{bm}
\newcommand{\beqy}{\begin{eqnarray}}
\newcommand{\eeqy}{\end{eqnarray}}
\newcommand{\mathbflet}{\begin{subequations}}
\newcommand{\emlet}{\end{subequations}}

\newcommand{\mm}{\mathrm}
\newcommand{\fm}{\,\text{fm}^{-3}}

\begin{document}

\def\gsimeq{\,\,\raise0.14em\hbox{$>$}\kern-0.76em\lower0.28em\hbox
{$\sim$}\,\,}
\def\lsimeq{\,\,\raise0.14em\hbox{$<$}\kern-0.76em\lower0.28em\hbox
{$\sim$}\,\,}

\title{Unified equations of state for cold nonaccreting neutron stars with Brussels-Montreal functionals. V. Improved parametrization of the nucleon density distributions}

\author{N.N. Shchechilin}
\affiliation{Institut d'Astronomie et d'Astrophysique, CP-226, Universit\'e
	Libre de Bruxelles, 1050 Brussels, Belgium}

\author{N. Chamel}
\affiliation{Institut d'Astronomie et d'Astrophysique, CP-226, Universit\'e
	Libre de Bruxelles, 1050 Brussels, Belgium}

\author{J.M. Pearson}
\affiliation{D\'ept. de Physique, Universit\'e de Montr\'eal, Montr\'eal
(Qu\'ebec), H3C 3J7 Canada}

\author{A.I. Chugunov}
\affiliation{Ioffe Institute, Politeknicheskaya 26, 194064 Saint-Petersburg, Russia}

\author{A.Y. Potekhin}
\affiliation{Ioffe Institute, Politeknicheskaya 26, 194064 Saint-Petersburg, Russia}

\date{\today}

\begin{abstract}
We previously studied the inner crust and the pasta mantle of a neutron star 
within the 4th-order extended Thomas-Fermi (ETF) approach with consistent proton 
shell corrections added perturbatively via the Strutinsky integral (SI) theorem 
together with the contribution due to pairing. To speed up the computations and 
avoid numerical problems, we adopted parametrized nucleon density distributions. 
However, the errors incurred by the choice of the parametrization are expected to become 
more significant as the mean baryon number density is increased, especially in the 
pasta mantle where the differences in the energy per nucleon of the different phases 
are very small, typically a few keV. To improve the description of these exotic 
structures, we discuss the important features that a nuclear profile should fulfill 
and introduce two new parametrizations. Performing calculations using the BSk24 
functional, we find that these parametrizations lead to lower ETF energy solutions for all pasta phases
than the parametrization we adopted before
and more accurately reproduce the exact equilibrium nucleon 
density distributions obtained from unconstrained variational calculations. 
Within the ETFSI method, all parametrizations predict the same composition 
in the region with quasi-spherical clusters. However, the two new parametrizations 
lead to a different mantle structure at mean baryon densities above about $0.07$~fm$^{-3}$, 
at which point lasagna is energetically favored. 
Interestingly, spherical clusters reappear in the pasta region. The inverted pasta phases 
such as bucatini and Swiss cheese are still found in the densest region above the core in all cases. 
\end{abstract}

\maketitle

\section{Introduction}
\label{intro}

After its birth in a gravitational core-collapse supernova explosion, a neutron star (NS) rapidly cools down.
After several weeks~\cite{PV19}, the region with  
mean baryon number densities $\bar n$ below about half-saturation density 
$n_0$ crystallizes. 
This solid crust consists of a crystal lattice of spherical or quasi-spherical 
neutron-rich nuclei 
embedded in a relativistic electron gas coexisting with free neutrons in its inner part. The core 
of the star remains a homogeneous liquid made of nucleons and leptons, and possibly other particles in 
the densest part (see, e.g., Ref.~\cite{blaschke2018} for a recent review). At the interface between 
the crust and the core, peculiar structures collectively referred to as `nuclear pasta' could 
emerge~\cite{Hashimoto+84,Ravenhall_ea83}. The existence of such a mantle would affect 
neutrino emission (see, e.g., Refs.~\cite{Gusakov_ea04_PastaURCA,Lin_ea20_PastaURCA}), electron transport 
important for thermal and electrical conductivities (see, e.g., Refs.~\cite{Yakovlev15_TraspPasta,pelicer2023}),
bulk viscosity~\cite{Yakovlev+17} and elastic properties~\cite{PP98,csh18_elast,Pethick+20_elst_past,Zemlyakov+23} 
of dense matter. This nuclear-pasta mantle could therefore be relevant to various NS phenomena. 

This paper is a continuation of our recent work on the structure of the 
nuclear-pasta mantle~\cite{Pearson+20,Pearson+22,Shchechilin+sym23}, in which we extended  
a semi-microscopic treatment  originally developed for ordinary 
nuclei~\cite{Dutta+86} and later adapted to the inner crust of a NS~\cite{Dutta2004,Onsi+08,Pearson+12,Pearson_ea15_pairing,Pearson_ea18_bsk22-26}. Our framework is based on the ETFSI approach, a computationally very fast 
approximation to the 
Hartree-Fock-Bogoliubov (HFB) calculations using semi-local functionals, such as those constructed from effective 
interactions of the Skyrme type. It is a two-stage method in which one first performs extended Thomas-Fermi (ETF)
calculations, followed by the addition of shell effects through the application
of the Strutinsky Integral (SI) theorem, and
the inclusion of pairing either via the Bardeen-Cooper-Schrieffer (BCS) method
or the local-density approximation. Both these corrections are made in a manner consistent with 
the ETF first stage. 

The ETF method \cite{Kirzhnits67,Grammaticos_Voros79,Grammaticos_Voros80,Brack_ea85} gives an approximation to the kinetic-energy density 
$\tau_q(\pmb{r})$ and the spin current $\pmb{J_q}(\pmb{r})$ in the functional.
It consists in expanding the Bloch density matrix in powers of $\hbar$ \cite{Wigner32,Kirkwood33}, thereby enabling $\tau_q(\pmb{r})$ and $\pmb{J_q}(\pmb{r})$ 
to be expressed as functions entirely 
of the nucleon densities $n_q(\pmb{r})$ and their gradients~\cite{Brack_ea85} (the highest 
degree of the derivatives corresponds to the order of expansion); the fourth-order terms are 
necessary for an accurate reproduction of the nuclear binding energies~\cite{Pi&Vinas+88,Centelles+90}. 
In this way, the energy 
becomes a functional of only the nucleon density distributions, and it is these, rather than wavefunctions, that are treated as the basic variables in a full Euler-Lagrange (EL) minimization for a given mean baryon number density $\bar{n}$. It was shown that the nucleon densities $n_q(\pmb{r})$ that satisfy the EL equations are smooth functions~\cite{barkat1972}. Assuming that pasta forms periodic structures, it
is enough to calculate the energy of a single Wigner-Seitz (WS) cell of the 
corresponding crystal lattice with a given number $Z$ of protons and $N$ of 
neutrons ($Z$ and $N$ represent the numbers per unit length in the case of
spaghetti and bucatini, and the numbers per unit area for lasagna). 
Such a cell is a truncated octahedron for a body-centered cubic 
crystal of quasi-spherical clusters or bubbles (Swiss cheese), an infinitely 
long regular hexagonal prism for a hexagonal lattice of  spaghetti or 
bucatini, and an infinitely long right prism for lasagna. The EL equations 
have to be solved with periodic boundary conditions, a consequence of which 
is that the gradient of the nucleon densities along directions perpendicular 
to the WS cell faces must vanish ~\cite{WignerSeitz33}. 

A full EL minimization is challenging, particularly for the three-dimensional 
case of quasi-spherical clusters and bubbles, and also for the
two-dimensional case of spaghetti or bucatini, since in both cases the
EL equations are partial differential equations. For this reason, the WS cell is generally approximated 
by more symmetrical cells of equal mean nucleon densities: a spherically symmetric sphere in the former case (this approximation was originally 
introduced by Wigner and Seitz in the context of electrons in 
solids~\cite{WignerSeitz33}), and an axially symmetric cylinder in the latter 
case. For lasagna, no approximation is needed, except that we assume it to be invariant under space inversion, thus ensuring a  consistent treatment of all geometries. 
Then, the smoothness requirement of the nucleon distributions together with the
periodicity lead to the boundary conditions
\beqy\label{eq:BC}
\dfrac{dn_q}{d\xi}(\xi=0)=
\dfrac{dn_q}{d\xi}(\xi=R)=0 \quad,
\eeqy
where $\xi$ is the radial coordinate for the spherical and cylindrical cells of radius $R$ and the Cartesian coordinate perpendicular to the plates of half-size $R$ in case of lasagna.

Even though the EL method has been reduced to a one-dimensional problem, 
solving the ordinary differential equations that have replaced the partial 
differential equations can still cause significant numerical problems, 
especially with the full implementation of the ETF method up to the fourth 
order~\cite{Brack_ea85}. In all our previous applications of the ETFSI method, the calculations were greatly simplified and computationally speeded up by parametrizing the nucleon density distributions. Without any loss of generality, the nucleon
profiles can be conveniently expressed as 
\beqy\label{eq:nq}
{n_q}(\xi;\mathbf{x}) = n_{\mm{B}q} + n_{\Lambda q}f_q(\xi;\bm{\chi})  \quad .
\eeqy
The parameter set $\mathbf{x}$ thus includes the background density $n_{\mm{B}q}$ and the parameter $n_{\Lambda q}$ that modulates the density excess or deficit due to the presence of clusters or holes respectively. 
The remaining parameters $\bm{\chi}$ are contained in the smooth dimensionless 
function $f_q(\xi;\bm{\chi})$ describing the spatial inhomogeneities. Once the form
of this function has been chosen the parameter set $\mathbf{x}$ for the given 
values of $N$ and $Z$ is determined 
by minimizing the total ETF energy per nucleon, which includes not only the nuclear energy but also the energy of the neutralizing electron gas and of the Coulomb
lattice~\cite{Pearson_ea18_bsk22-26,Pearson+20}.

With the ETF part of the ETFSI calculation completed, microscopic corrections 
due to proton shell effects are added, as described in Ref.~\cite{Pearson+22}. 
The analogous corrections for neutrons are expected to be negligible because of 
the occupation of (quasi)continuum states~\cite{Oyamatsu&Yamada94,Chamel06,Chamel+07}. 
For consistency, we drop the shell correction when the Fermi energy exceeds 
the value of the potential at the border of the cell.  Here we ignore proton pairing 
since its impact was shown to be quite marginal in the pasta mantle~\cite{Pearson+22,Shchechilin+sym23}. 
Neutron pairing effects will be the subject of a future study. 

The main concern of this paper is with the choice of a smooth profile
function $f_q(\xi;\bm{\chi})$.
In the original application of the ETFSI method to finite nuclei (for which 
$n_{\mm{B}q}=0$), the simple Fermi-Dirac distribution was adopted~\cite{Dutta+86},
\beqy\label{eq:fd}
f^\mm{FD}_q(\xi;C_q,a_q) = \frac{1}{1 +\exp \Big(\frac{\xi-C_q}{a_q}\Big) } \quad ,
\eeqy
where $C_q$ is the half-width nuclear radius and $a_q$ accounts for the diffuseness of the 
the nuclear surface. This form has also been recently employed for ETFSI 
calculations of the inner crust of a NS~\cite{ShelleyPastore2021}. However, the 
parametrization~\eqref{eq:fd}  (like the majority of those employed for ordinary nuclei, e.g. Refs.~ \cite{Bethe68,Brueckner+69,Moszkowski70,Blaizot&Grammaticos81,Kolehmainen_ea85}) 
does not appear to be very well suited in this context since it does not satisfy 
the boundary conditions~\eqref{eq:BC}. The associated errors can be significant, especially 
in the pasta region where the size of the clusters or holes is comparable to the distance 
between them. 

For this reason, the following modification was proposed in Ref.~\cite{Onsi+08} 
and adopted in all our subsequent studies including Refs.~\cite{Pearson+20,Pearson+22,Shchechilin+sym23}: 
\beqy\label{eq:strd}
f^\mm{StrD}_q(\xi;C_q,a_q) = \frac{1}{1 +\exp \left[\Big(\frac{C_q - R}
{\xi - R}\Big)^2 - 1\right]\exp \Big(\frac{\xi-C_q}{a_q}\Big) } \quad . 
\eeqy
The additional exponential factor ensures that not only the first derivative 
of ${n_q}(\xi;\mathbf{x})$ vanishes at the border of the cell but in fact that 
\emph{all} derivatives do so. The vanishing of the first three derivatives makes
possible a substantial simplification of the 4th-order ETF expressions for the
energy by integrating by parts~\cite{Brack_ea85,Onsi+97}. However, the
vanishing of all derivatives on the cell surface leads inevitably to a
strong damping with a very flat tail. Moreover, this profile, like the 
Fermi profile (\ref{eq:fd}), has the further defect of a kink at the origin,
i.e., a non-zero first derivative. Evaluating the impact of these 
limitations in the parametrization~\eqref{eq:strd} and investigating the 
necessary improvements constitute the objective of this paper. 

In Section~\ref{subsec:choice}, we introduce two new parametrizations for the 
nucleon density distributions and we show in Section~\ref{subsec:res_etf} that 
they lead to a more accurate description of nuclear pasta by performing ETF 
calculations with functional BSk24~\cite{Goriely_ea_Bsk22-26}. Complete ETFSI calculations with these new 
profile functions are presented and discussed in Section~\ref{subsec:res_si}. 
Our conclusions are given in Section~\ref{sec:concl}.

\section{Optimization of nuclear profiles}\label{sec:optim}

\subsection{Choice of profiles}\label{subsec:choice}

\begin{figure}	\includegraphics[width=0.7\columnwidth]{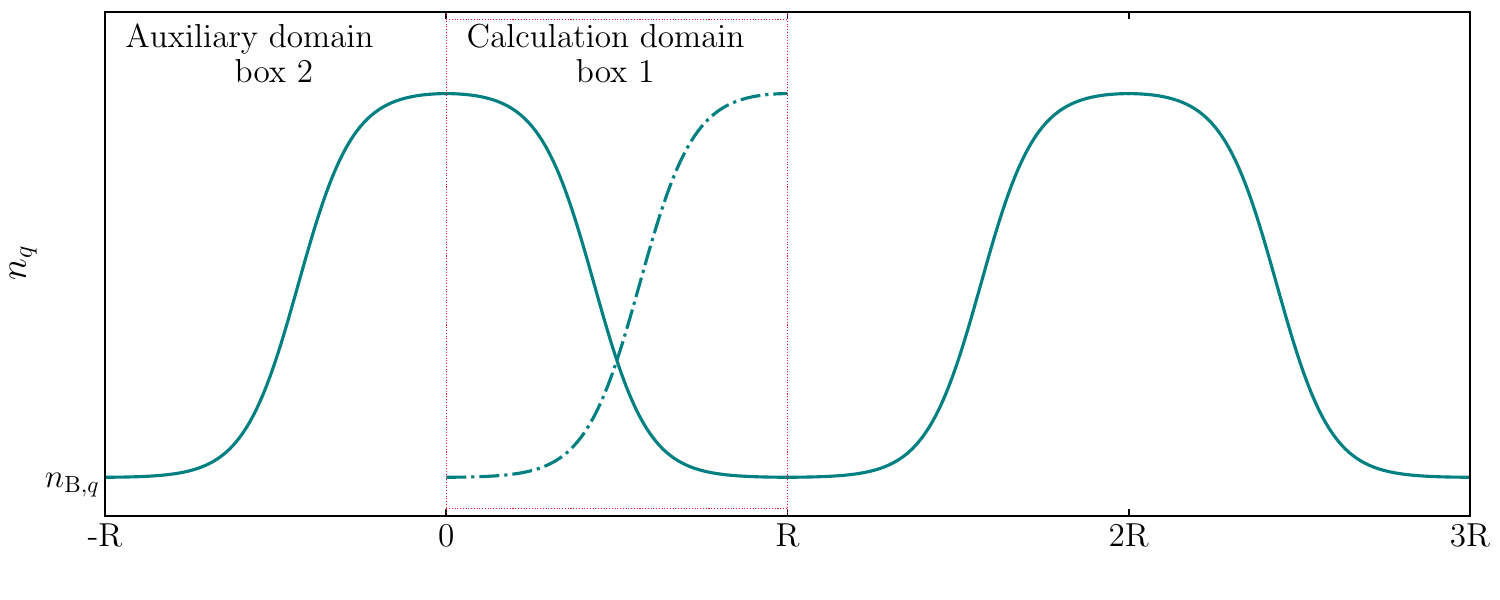}
	\caption{Illustration of the identity between the seemingly different lasagna (green solid line) 
 and anti-lasagna  (green dash-dotted line) configurations in the calculation domain (dotted red box). 
  These configurations are actually not physically distinct since the anti-lasagna profile can be 
  obtained from a space inversion followed by a translation of the lasagna profile. This symmetry
  leads to the condition~\eqref{eq:lasagna_req}.}
\label{fig:example}
\end{figure}

\begin{figure}	\includegraphics[width=0.38\columnwidth]{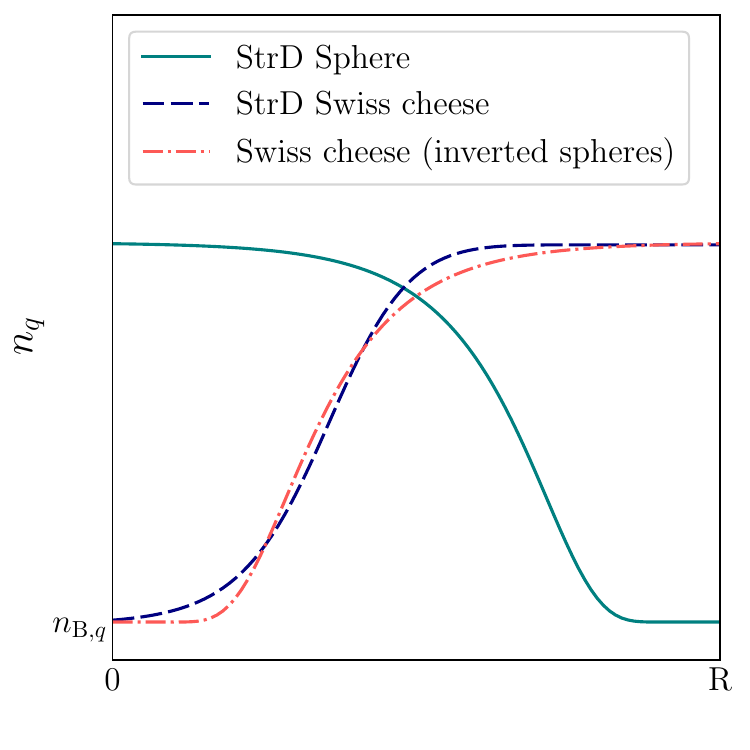}
	\caption{Schematic picture illustrating possible nucleon distributions inside the WS cell using the StrD parametrization~\eqref{eq:nq} and \eqref{eq:strd}:
   `flat-bottom' spheres (green solid line) and `flat-top' Swiss cheese (blue dashed line). The `flat-bottom' Swiss cheese profile (red dash-dotted line) obtained from the inversion of spheres
   cannot be described by this parametrization (see text for discussion). }
\label{fig:example2}
\end{figure}

Besides being a smooth function, satisfying the boundary conditions \eqref{eq:BC}, the parametrization should be flexible enough to describe not only 
phases with clusters immersed in a more dilute medium but also inverted phases 
such as bucatini and Swiss cheese. The case of lasagna is peculiar in that the distinction between these two different topologies vanishes: although a configuration 
with a `hole' at the center of the cell appears as an anti-lasagna, it can be 
transformed into a lasagna by a simple translation, as illustrated by the green dash-dotted 
line in Fig.~\ref{fig:example}. In other words, these two seemingly different configurations 
are physically identical. To avoid introducing a spurious distinction between lasagna and 
anti-lasagna, both should be describable equally well by the chosen parametrization. This 
means that for any given profile $n_q(\xi;\bf{x})$, there should exist a set of parameters 
$\bf{x^\prime}$ such that the inverted
profile ${n_q}(R-\xi;\mathbf{x^\prime})$ obtained from a space inversion followed by a 
translation is also allowed, as shown  
in Fig.~\ref{fig:example}: 
\beqy\label{eq:lasagna_req}
{n_q}(\xi;\mathbf{x})={n_q}(R-\xi;\mathbf{x^\prime}) \, .
\eeqy 
This condition guarantees that these lasagna and anti-lasagna configurations 
have the same energy.
Even though this argument is not applicable in the case of spheres and
cylinders because of the WS approximation, there are still reasons why 
Eq.~\eqref{eq:lasagna_req} would be desirable in those cases. Indeed, 
the condition~\eqref{eq:lasagna_req} ensures equal treatment of 
spheres (spaghetti) and Swiss cheese (bucatini) since for any profile 
obtained for the former, the inverted one will be also allowed for the 
latter. The strong damping parametrization does not satisfy this requirement, 
thus leading to an artificial anti-lasagna configuration. Moreover, it 
can only describe `flat-bottom' spheres and `flat-top' 
bubbles, as illustrated in Fig.~\ref{fig:example2} by the solid green and 
dashed blue lines respectively. But `flat-bottom' Swiss cheese (corresponding 
to the red dash-dotted line) are not allowed. The same limitation 
applies to the spaghetti and bucatini shapes.

To improve the description of pasta and assess the sensitivity of our results with respect to the choice of the parametrization, we first consider the `3FD' 
parametrization consisting of a combination of three Fermi-Dirac 
functions~\eqref{eq:fd} as follows:
\beqy\label{eq:3fd}
f^\mm{3FD}_q(\xi;C_q,a_q) = f^\mm{FD}_q(-\xi;C_q,a_q) + f^\mm{FD}_q(\xi;C_q,a_q)+f^\mm{FD}_q(2R-\xi;C_q,a_q)-f^\mm{FD}_q(-R;C_q,a_q)-2f^\mm{FD}_q(R;C_q,a_q)\quad ,
\eeqy
where the constant are subtracted so that $f^\mm{3FD}_q(R;C_q,a_q)=0$. 
Raising the Fermi-Dirac distributions to the power $\gamma$ (see, e.g.,  Refs.~\cite{Kolehmainen_ea85,Brack_ea85}) does not 
lead to a substantial energy reduction in the pasta region (as was also found in 
Ref.~\cite{MU15}), therefore we set $\gamma=1$
here. Apart from all the advantages of
the simple Fermi-Dirac form\footnote{Actually, the results obtained within the fourth-order ETF method for FD and 3FD 
parametrizations are almost indistinguishable at mean baryon densities below $\bar{n}\approx 0.05\fm$. However, the FD
parametrization becomes numerically unstable at higher densities due to increasingly large values of the derivatives
of $n_q$ at the surface of the integration volume.  
Incidentally, 
we check that our results within the second-order ETF method with the 3FD parametrization using the SLy4  
functional~\cite{Chabanat_ea98_SLY4} are in reasonable agreement with those of Ref.~\cite{MU15}.}, e.g. fulfilling the
condition~\eqref{eq:lasagna_req} with $n_{\Lambda q}^\prime=-n_{\Lambda q}$, $a_q^\prime=a_q$, $C_q^\prime=R-C_q$, and
$n_{{\rm B}q}^\prime=n_{{\rm B}q}+n_{\Lambda q}(3-f^\mm{FD}_q(-R;C_q,a_q)-2f^\mm{FD}_q(R;C_q,a_q)-f^\mm{FD}_q(-R;C_q^\prime,a_q^\prime)-2f^\mm{FD}_q(R;C_q^\prime,a_q^\prime))$, the additional merit is that this
parametrization is smoother although the first derivatives at $\xi=0$ and $\xi=R$ are still not strictly zero\footnote{Complementing the 3FD parametrization with two additional Fermi-Dirac functions (suppressing the 
derivatives even more) does not change the results.}.

On the other hand the satisfaction of the boundary condition \eqref{eq:BC} can 
be exactly achieved by generalizing the strong-damping 
expression~\eqref{eq:strd} to 
\beqy\label{eq:strd_gen}
f^\mm{}_q(\xi;C_q,a_q) = \frac{1}{1 + 
h(\xi;C_q,a_q) \exp \Big(\frac{\xi-C_q}{a_q}\Big) } \quad ,
\eeqy
in which the function $h(\xi;C_q,a_q)$ ensures the fulfillment of the conditions \eqref{eq:BC}.
Moreover, if the function is such that 
\beqy\label{eq:lasagna_req_softd}
h(\xi;C_q,a_q) = \frac{1}{h(R - \xi; R-C_q, a_q)} \quad ,
\eeqy
it follows that the constraint~\eqref{eq:lasagna_req} can be easily satisfied with
$n_{\Lambda q}^\prime=-n_{\Lambda q}$, $n_{{\rm B}q}^\prime =n_{{\rm B}q}+n_{\Lambda q}$, 
$a_q^\prime=a_q$, and $C_q^\prime=R-C_q$. 
The simplest function $h(\xi;C_q,a_q)$ that we have found obeying these requirements is: 
\beqy\label{eq:softd}
h(\xi;C_q,a_q) = 
\left(\frac{C_q-R} {C_q}\right)^2 \left(\frac{\xi}{\xi-R}\right)^2 \quad .
\eeqy
Only the first derivative of this new profile, which we will refer to as `SoftD' below, 
vanishes on the cell surface. This SoftD profile therefore allows for a softer damping than
the StrD profile~\eqref{eq:strd}. 

\subsection{Results of ETF minimization}\label{subsec:res_etf}

To optimize the minimization of the ETF energy, we start the calculations in the shallowest layers of the inner crust containing quasi-spherical nuclei for which fairly accurate initial guesses for the parameters of the profiles can be set: $C_q \approx 6$~fm,  $a_q\approx0.5$~fm, $n_{\Lambda n}\approx N/(\frac{4}{3}\pi C_n^3)$ and $n_{\Lambda p}\approx Z/(\frac{4}{3}\pi C_p^3)$. The final values are then taken as initial guesses for the deeper neighboring layer. The process is repeated until the crust-core boundary is reached. We have checked the robustness of our results by varying the initial guesses at different densities. It is important to scan regions of parameter space large enough to ensure that our minima are true minima minimora, and not just local minima. 

We now compare the results obtained with the three profile parametrizations. The StrD profile~\eqref{eq:strd}, 
which we implemented in our previous works and, as was mentioned earlier, involves an integrated version of the fourth-order ETF method
~\cite{Brack_ea85,Onsi+97}, is used as a reference. In addition, we consider the two new profiles 3FD~\eqref{eq:3fd} and SoftD~\eqref{eq:strd_gen}, \eqref{eq:softd} for which we use full ETF expressions for $\tau_q(\pmb{r})$ and $\pmb{J_q}(\pmb{r})$ given in Refs.~\cite{Grammaticos_Voros79,Grammaticos_Voros80,Bartel+02}, since only the first derivatives at $\xi=0$ and $\xi=R$ vanish. In this case, the only price to pay is a slight increase in the computational time. 

\begin{figure}
	\includegraphics[width=0.7\columnwidth]{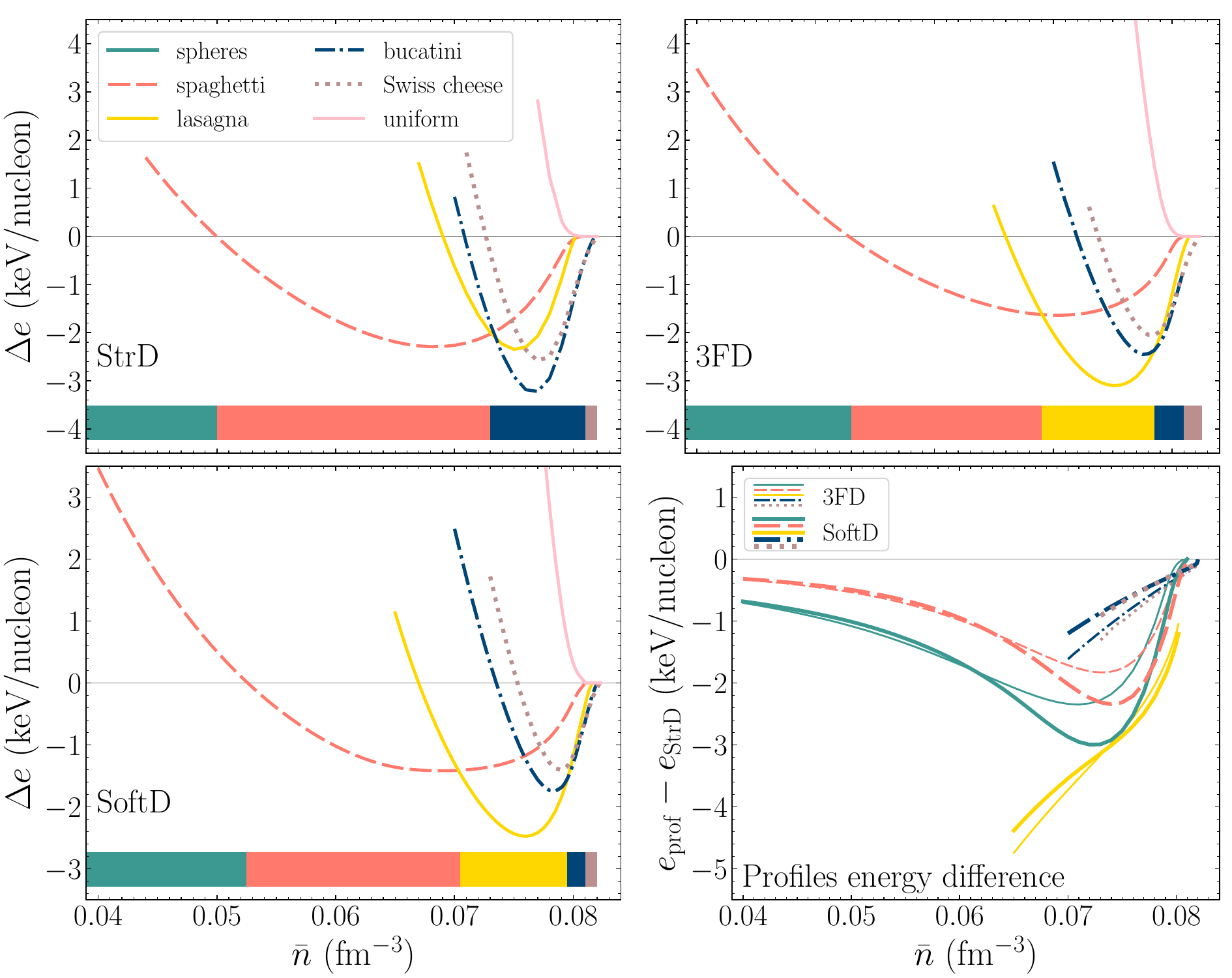}
	\caption{(upper panels and lower left panel) Energy per nucleon of different pasta phases minus the energy per nucleon of spheres within the ETF approach versus the mean baryon density for the three adopted parametrizations of the nucleon profiles. Orange dashed lines correspond to spaghetti, solid yellow to lasagna, navy dash-dotted to bucatini, brown dotted to Swiss cheese, and solid pink to uniform $npe$ matter. The equilibrium shape is illustrated by the bars at the bottom of the panels with the same color coding. (lower right panel) The energy difference between the two new profiles and StrD~\cite{Onsi+08} for various pasta shapes. Solid green lines are added for spheres. The thin lines represent the 3FD profile and the thick ones are for the SoftD, while colors and line styles are the same as in the two upper panels and lower left one.}
	\label{fig:etf}
\end{figure}

Since the ETF approach is completely variational, it is this method that should serve as a basis for choosing the optimal parametrization. To this end, we minimize the energy per nucleon for each of the three parametrizations considering the five different phases: spheres, spaghetti, bucatini, lasagna, and Swiss cheese. Figure~\ref{fig:etf} shows the corresponding energy 
per nucleon with the energy for spheres subtracted. The corresponding equilibrium pasta configurations are indicated by the color bar at the bottom of each panel. The results for the StrD are identical to the ones presented in Ref.~\cite{Shchechilin+sym23}. 
The most remarkable difference obtained with the two new profiles is the reappearance of lasagna between the spaghetti and bucatini phases. 
The onset of the pasta phases is slightly shifted from $\bar{n}_\mathrm{sp}\approx0.05\fm$ for StrD to $\bar{n}_\mathrm{sp}\approx0.052\fm-0.053\fm$ for the two new profiles. The crust-core transition density is barely affected, being $\bar{n}_\mathrm{cc}\approx0.082\fm$. It corresponds to the point, for which the energy of the densest pasta phase, namely the Swiss cheese for all profiles, is equal to the energy of uniform $npe$ matter (characterized by $n_{\mm{\Lambda}q}=0$).
Within the same nuclear functional, the compressible liquid drop model can yield a similar pasta sequence, but only if the curvature correction is included \cite{DinhThi+21a,Zemlyakov+21}.

To better understand these results, we demonstrate in the bottom right panel of Fig.~\ref{fig:etf} the energy per nucleon of each phase obtained with the two new parametrizations after subtracting out the energy per nucleon for StrD. It is clear that in all cases the new parametrizations yield lower energies. In general, SoftD performs slightly better than 3FD for the clusterized phases, while 3FD
is preferred for the inverted phases. The energy differences amount to a maximum of about $\approx2-3$~keV per nucleon for spheres and spaghetti 
and is reached near $\bar{n}\approx0.074\fm$. For the inverted phases at the relevant densities, the energy reduction with the new 
parametrizations does not exceed $\approx1$~keV per nucleon. The largest deviations of order $\lesssim4-5$~keV per nucleon are found in the lasagna phase and 
are enough to change the sequence of pasta. 

\begin{figure}
	\includegraphics[width=\columnwidth]{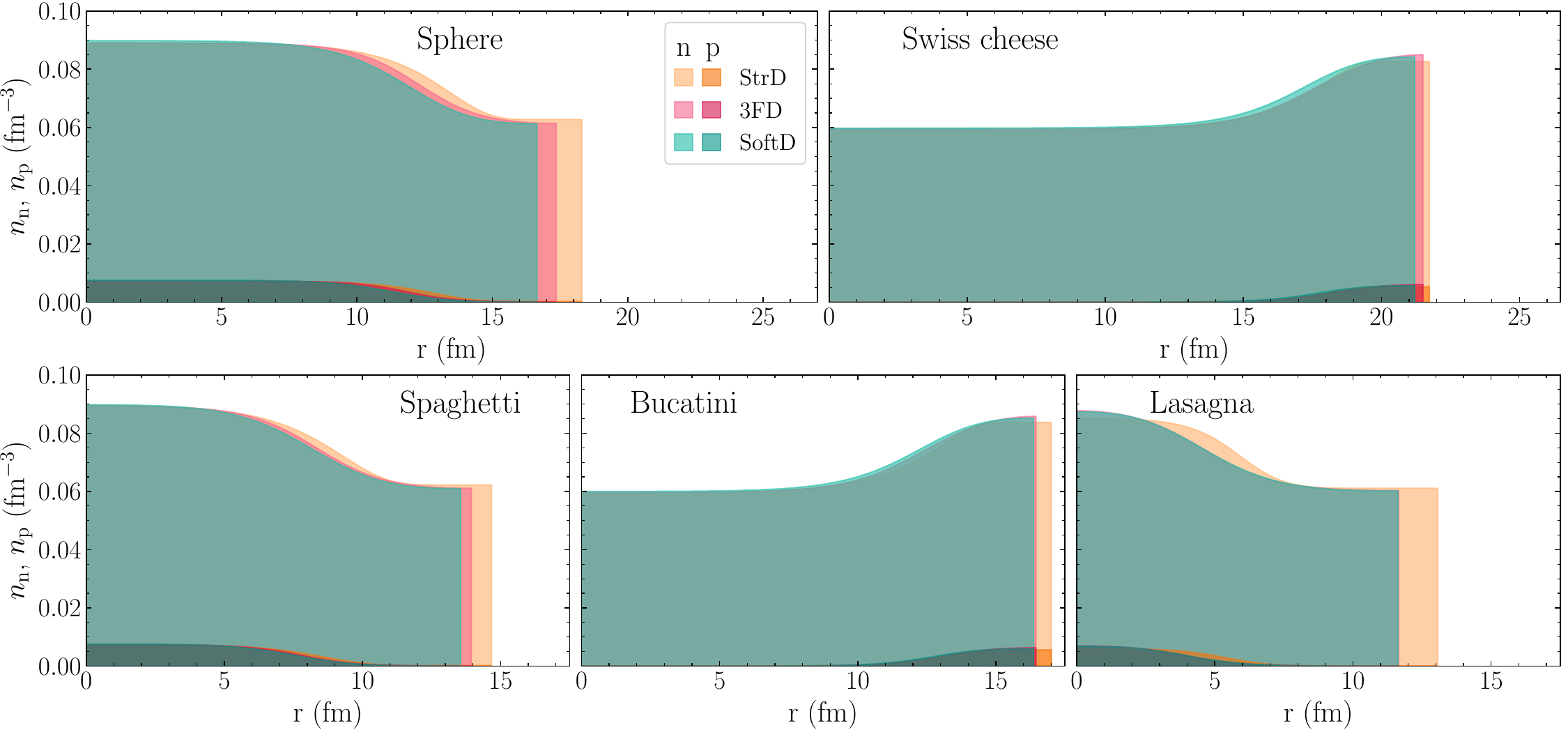}
	\caption{Equilibrium nucleon density profiles for three parametrizations at $\bar{n}=0.074\fm$. Green shading corresponds to SoftD, pink for 3FD, and orange for StrD. Proton distributions are plotted with darker colors than neutrons.}
	\label{fig:etf_prof}
\end{figure}

As an illustrative example, we plot in Fig.~\ref{fig:etf_prof} the equilibrium nucleon distributions in the WS cell at $\bar{n}=0.074\fm$ for the three different parameterizations. The two new SoftD and 3FD profiles vary more smoothly than StrD 
and lead to smaller cells. The difference is again more prominent for the lasagna phase, complementing the conclusions of Fig.~\ref{fig:etf}. In addition, we show in Appendix~\ref{appB} that the two new profiles (unlike StrD) are flexible enough to reproduce the EL solutions in the TF approximation obtained in Ref.~\cite{Sharma_Centelles+15}. This suggests that 3FD and 
SoftD distributions lie very close to the exact profiles and are preferred to the StrD.

\section{Results of full ETFSI calculations}\label{subsec:res_si}

At the next step following the approach of Refs.\cite{Pearson+22,Shchechilin+sym23}, we add the SI corrections on top of the ETF energy per nucleon for each of the three profile parametrizations StrD, 3FD, and SoftD.  In the low-density layers of the inner crust, clusters are quasi-spherical and sufficiently far apart so that the differences in the profiles at the ETF level turn out to be insignificant (see also Fig.~\ref{fig:etf}). Moreover, the proton shell effects are more pronounced leading to deeper local minima in the ETFSI energy per nucleon. Consequently, all the profile parametrizations predict the same
composition.

\begin{figure}
	\includegraphics[width=\columnwidth]{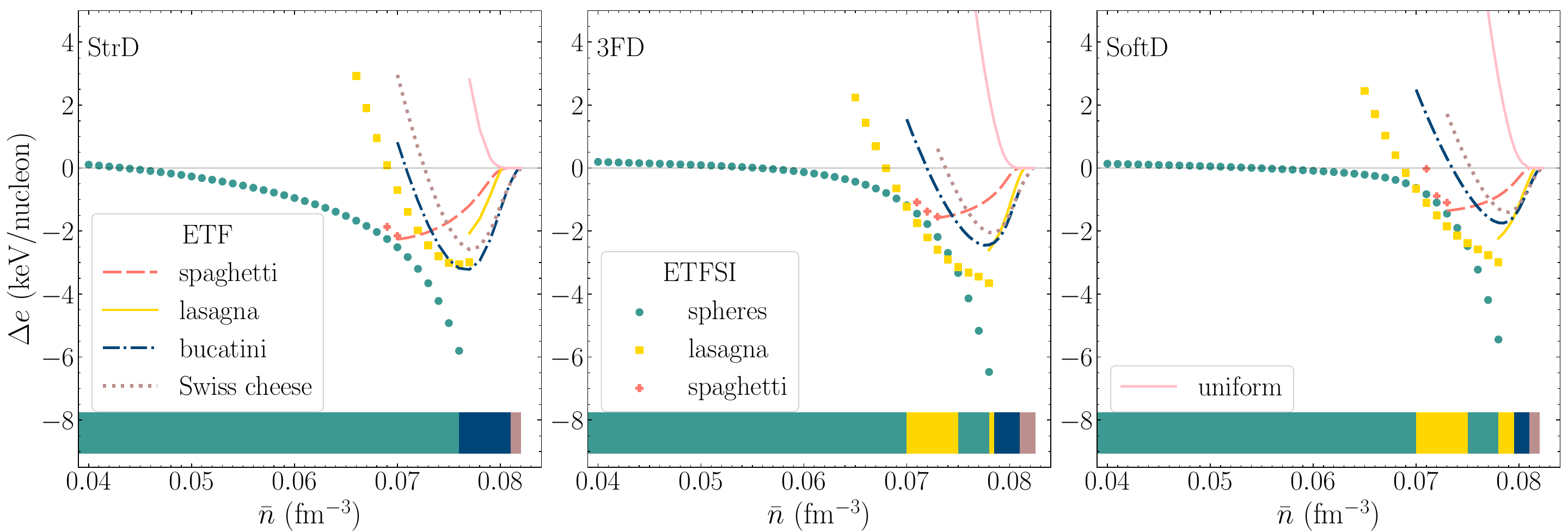}
	\caption{ETFSI and ETF energy per nucleon for different pasta phases minus the ETF energy per nucleon for spheres versus mean baryon density for the three considered profile parametrizations. The ETFSI results for spheres (solid green circles) and lasagna (solid yellow squares) are supplemented with ETF results (the same as in Fig.~\ref{fig:etf}) beyond proton drip for the corresponding phase.  For spaghetti, all configurations shown (solid pluses - ETFSI, dashed lines - ETF) are beyond proton drip (see text for details). For completeness, the ETF energies per nucleon for bucatini and Swiss cheese and the results for the homogeneous nuclear matter are plotted (see also Fig.~\ref{fig:etf}). Color bars at the bottom indicate the equilibrium phases.}
	\label{fig:etfsi}
\end{figure}

However, in the deeper region of the pasta mantle at densities above $\approx 0.06$~fm$^{-3}$, the choice of the parametrization matters. In Fig.~\ref{fig:etfsi}, we display the ETFSI results over the whole density range of pasta phases for the three adopted 
parametrizations. As found earlier in Refs.~\cite{Pearson+22,Shchechilin+sym23}, the spaghetti phase predicted at the ETF level 
totally disappears when the SI correction is added for all three parametrizations (only the configurations with dripped protons appear in the
scale of Fig.~\ref{fig:etfsi} beyond $\bar{n}\approx0.067\fm$). As a consequence, the spherical clusters remain present at higher densities
thus shrinking the pasta mantle substantially. In contrast to the StrD, the SoftD, and the 3FD parametrizations lead to the presence of lasagna in the density range $\bar{n}_\mm{sp}\approx 0.07\fm-0.08\fm$. Interestingly, this layer occurs to be interspersed by spheres around $\bar{n}\approx 0.075\fm$. 
The existence of the same pasta phases in different regions of a NS was previously discussed  
in Ref.~\cite{Magierski&Heenen02}. Beyond $\bar{n}_\mm{drip}\approx0.078\fm$, some protons become unbound for all considered shapes, and since we then drop the SI corrections, the presence of bucatini and Swiss cheese as well as the crust-core transition are unaffected. As shown in  
Fig.~\ref{fig:etfsi}, the total ETFSI results for the two new parametrizations are quite the same. 

\begin{figure}
 	\includegraphics[width=0.9\columnwidth]{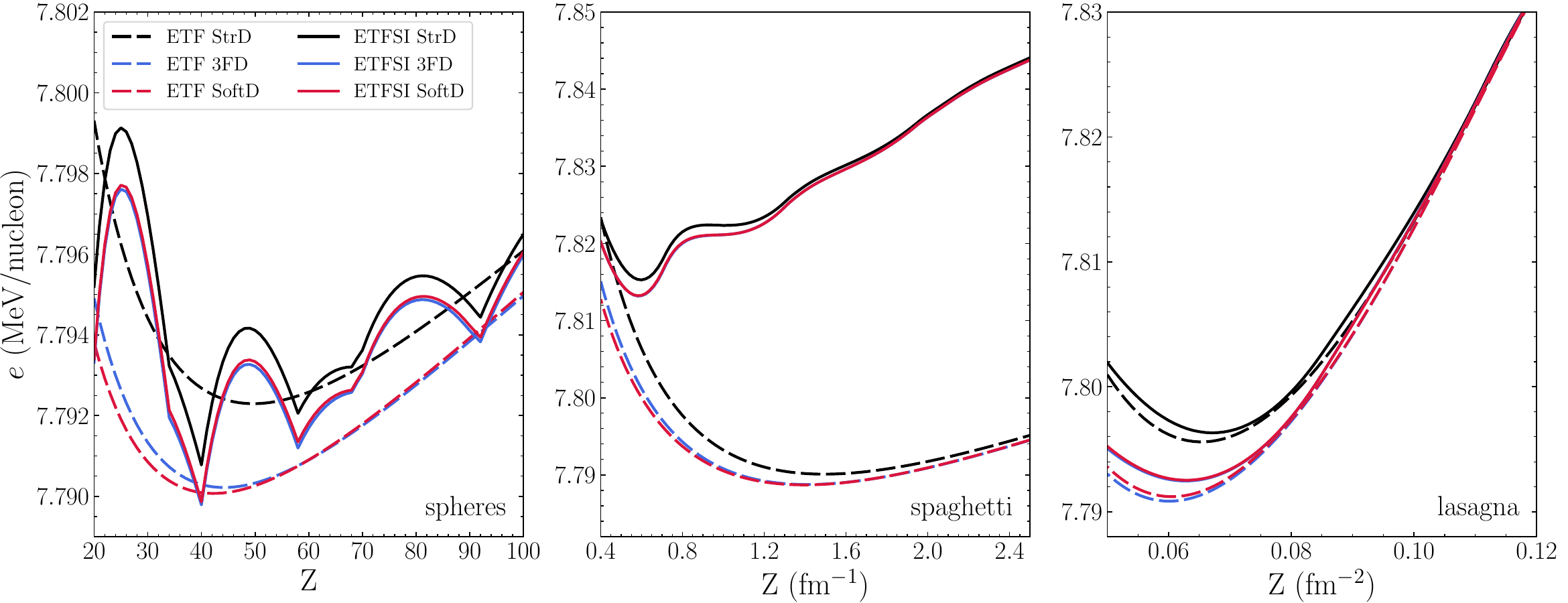}
 	\caption{ETF (dashed lines) and ETFSI (solid lines) energies per nucleon versus proton number for spheres, proton number per unit length for spaghetti and proton number per unit area for lasagna at baryon density $\bar{n}=0.065\fm$. Three different parametrizations of the nucleon density distributions were considered: 3FD (blue lines), SoftD (red lines), StrD (black lines).}
 	\label{fig:esi65}
 \end{figure}

To analyse in more detail the role of the initial ETF parametrizations, we fix the density to $\bar{n}=0.065\fm$ and plot the energies per particle as a function of $Z$ in Fig.~\ref{fig:esi65}. For spheres, a typical shell structure is found for
all parametrizations with a sharp minimum at $Z=40$. In the case of spaghetti and lasagna, the energy per nucleon also exhibits some smooth
fluctuations with $Z$ associated with the filling of levels (see also Ref.~\cite{Oyamatsu&Yamada94}). 
In Ref.~\cite{Pearson+22}, the interval of the $Z$ considered was too small to reveal these fluctuations. The SI correction for spaghetti
leads to a substantial reduction of $Z$ by more than a factor of 2, from $Z_\mm{eq}\approx 1.4$\,fm$^{-1}$ to $Z_\mm{eq}\approx 0.6$\,fm$^{-1}$. For such small values of $Z_\mm{eq}$, the SI correction is reduced but at the cost of increasing the ETF energy. The 
end result is that the ETFSI energy per nucleon is larger than for spheres and lasagna (note the different scales in
Fig.~\ref{fig:esi65}) leading to the vanishing of spaghetti. As discussed in Appendix~\ref{appC}, this can be understood from 
the different filling of energy levels. 

Comparing the results obtained for the three different parametrizations, the 
StrD is found to yield the highest ETFSI energy per nucleon for all phases. As 
discussed before, the energy reduction with the new parametrizations has the biggest
impact on the lasagna phase. The SoftD and 3FD lead to remarkably close predictions. Notably, it can also be seen that differences in the energy per nucleon among the three parametrizations tend to be reduced when the SI correction is added. Nevertheless, differences still exist, especially between the StrD and the two new parametrizations. This is due to the perturbative nature of the corrections.  

 \label{subsec:profiles_si}

\begin{figure}
	\includegraphics[width=\columnwidth]{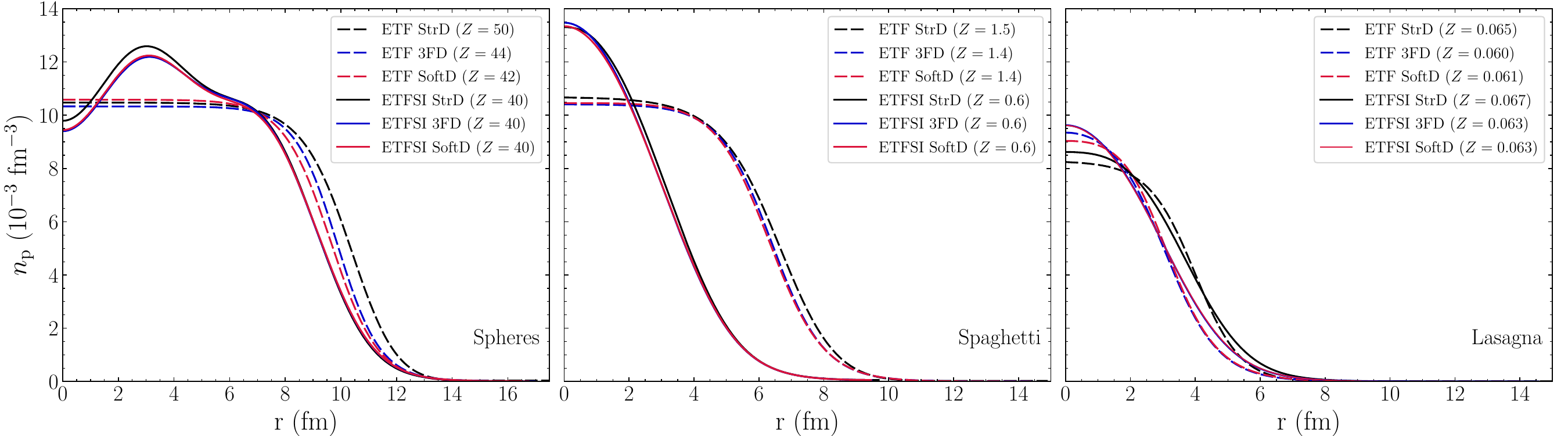}
	\caption{Proton density distributions profiles, as obtained within the ETFSI (solid lines) or ETF (dashed lines) method, for spheres, spaghetti, and lasagna at $\bar{n}=0.065\fm$. Red lines correspond to SoftD, blue for 3FD, and black for StrD. Each profile is plotted for its own optimum $Z$ value (see Fig.~\ref{fig:esi65}).}
	\label{fig:prof65}
\end{figure}

These results are also reflected in the proton density distributions plotted in Fig.~\ref{fig:prof65}. There the parametrized 
ETF profiles are compared to those calculated in the ETFSI approach (for the corresponding optimal $Z$ value that is different from 
that in the ETF calculation) by summing the occupied single-particle wave functions obtained from the solutions of the HF equations 
with the ETF mean fields.
The effect of the SI corrections is the most significant for spaghetti and can be traced back to the drop of $Z$. For spheres, the ETFSI treatment 
induces quantum fluctuations inside the cluster. For lasagna, the initial parametrized profiles can already be considered as a fairly
good approximation to the ETFSI profiles (see also Appendix~\ref{appB} for further comparisons).

\section{Conclusions}
\label{sec:concl}

We have determined the structure of the deepest layers of the inner crust and of the pasta mantle of a NS within the ETFSI approach, 
investigating the role of the parametrization of the nucleon distributions. To this end, we have formulated the requirements that 
the profiles should satisfy. (i) The first derivatives should vanish at the center and at the border of the WS cell, as expressed 
in Eq.~\eqref{eq:BC}. (ii) The profile should satisfy Eq.~\eqref{eq:lasagna_req} to avoid introducing an artificial distinction 
between lasagna and anti-lasagna. (iii) The parametrization should be flexible enough to reproduce as closely as possible the exact 
profiles obtained from the solutions of the EL equations. We have shown that the various parametrizations employed for isolated nuclei 
(in vacuum) \cite{Brueckner+69,Bethe68,Brack_ea85,Kolehmainen_ea85,Moszkowski70,Blaizot&Grammaticos81} and even the ones specifically
developed for the inner crust (including ours) \cite{Oyamatsu93,Gogelein&Muther07,MU15,Lim&Holt17,Mondal+20,Onsi+08,Pearson_ea18_bsk22-26} have several drawbacks limiting their 
reliability, especially in the pasta layers. For this reason, we have introduced the parametrizations SoftD and 3FD (see 
Eqs.~\eqref{eq:nq}, \eqref{eq:strd_gen}, \eqref{eq:softd}, and \eqref{eq:3fd}), fulfilling the aforementioned requirements.  

First, we have compared at the ETF level these two new parametrizations with the StrD parametrization we proposed in Ref.~\cite{Onsi+08}. 
For numerical calculations, we have applied the generalized Skyrme functional BSk24~\cite{Goriely_ea_Bsk22-26} as in our recent 
series of works~\cite{Pearson_ea18_bsk22-26,Pearson+20,Pearson+22,Shchechilin+sym23}. We have found that 
the Soft and 3FD parametrizations yield lower energies for all pasta phases, notably for lasagna. This 
shows that the new parametrizations are more realistic, leading to configurations that are more stable 
and therefore closer to the exact EL solutions. The obtained energy reductions are substantial to alter the equilibrium structure of the pasta mantle. Namely, in contrast to the StrD parametrization, the SoftD 
and 3FD parametrizations lead to the same traditional sequence of pasta with increasing density, as 
predicted by most compressible liquid drop models~: spheres, spaghetti, lasagna, bucatini, and Swiss cheese. 

When including the SI corrections, all three parametrizations agree in their predictions up to the point where pasta appears, at densities $\bar{n}_\mm{sp}\approx0.07\fm$. This transition density is significantly increased compared to the ETF results. This is caused by the vanishing of spaghetti as we previously found in 
Refs.~\cite{Pearson+22,Shchechilin+sym23} using the StrD parametrization. This stems from the fact that the SI correction is large and positive, which in turn arises because of a comparatively larger number of protons occupying the most loosely bound levels 
(see Fig.~\ref{fig:pot}). Despite the tendency of the SI correction to compensate 
the differences in the ETF energy from initially parametrized profiles, the StrD results differ from the two new parametrizations at densities above $\bar{n}_\mm{sp}\approx0.07\fm$. In this region, both the SoftD and 3FD parametrizations predict lasagna to be present, which is, 
surprisingly, interspersed among spherical clusters. In all cases, the 
densest layers of the pasta mantle consist of inverted phases, namely 
bucatini and Swiss cheese, and the transition to the homogeneous core occurs 
at $\bar{n}_\mm{cc}\approx0.082\fm$. 
The end result is that the pasta mantle calculated with the two new parameterizations has a more complicated structure and extends over a slightly wider range of densities than obtained with the StrD, though considerably reduced compared to ETF results.  
All in all, the new proposed nuclear profile parametrizations yield similar results, predicting a pasta mantle structure closer to the true equilibrium. These parameterizations are therefore more realistic and better suited for investigating the role of different nuclear functionals in the formation of nuclear pasta.

\begin{acknowledgments}
This work benefited from valuable discussions with Mikhail E. Gusakov to whom the authors are grateful.
N.N.S. and N.C. are thankful to Michael Urban and Francesca Gulminelli for their comments. The work of N.N.S. was 
financially supported by the FWO (Belgium) and the Fonds de la Recherche Scientifique (Belgium) under the Excellence of Science (EOS) programme (project No. 40007501). The work of N.N.S., N.C., and J.M.P. also received funding from the Fonds de la Recherche Scientifique (Belgium) under Grant No. IISN 4.4502.19.
\end{acknowledgments}

\appendix


\section{Flexibility of the profile parametrizations}
\label{appB}

Here we provide a simple test of the flexibility of the different parametrizations by trying to 
fit directly the parameters to realistic nuclear profiles in the NS mantle (without performing the 
ETF(SI) minimization). 
For this purpose, we have taken the nucleon density distributions obtained 
in Ref.~\cite{Sharma_Centelles+15} from the solution of the EL equations within the TF approximation at different mean baryon densities. We also include in the analysis the profiles from self-consistent HF calculations of the lasagna phase for the
fixed proton fraction $Y_\mm{p}=0.1$ and $\bar{n}=0.04$\,fm$^{-3}$~\cite{Sekizawa+22}. 
For the sake of completeness, in addition to the parametrizations StrD, 3FD, and SoftD, we have 
considered the parametrization widely applied for TF calculations~\cite{Oyamatsu93,Gogelein&Muther07,Miyatsu+13,Lim&Holt17} 
\beqy\label{eq:PW}
f^\mm{PW}_q(\xi;C_q,a_q) =\begin{cases} \left[1 -\left(\dfrac{\xi}{C_q}\right)^{a_q}\right]^{3}, & r<C_q,\\[5mm]
0, & r\geq C_q.
\end{cases} \eeqy 
Although the form~\eqref{eq:PW} satisfies the boundary conditions \eqref{eq:BC}, it leads to divergences for the 4-th order ETF energy. It is also clear 
that the condition~\eqref{eq:lasagna_req} cannot be fulfilled for $C_q<R$.  

\begin{figure}
	\includegraphics[width=\columnwidth]{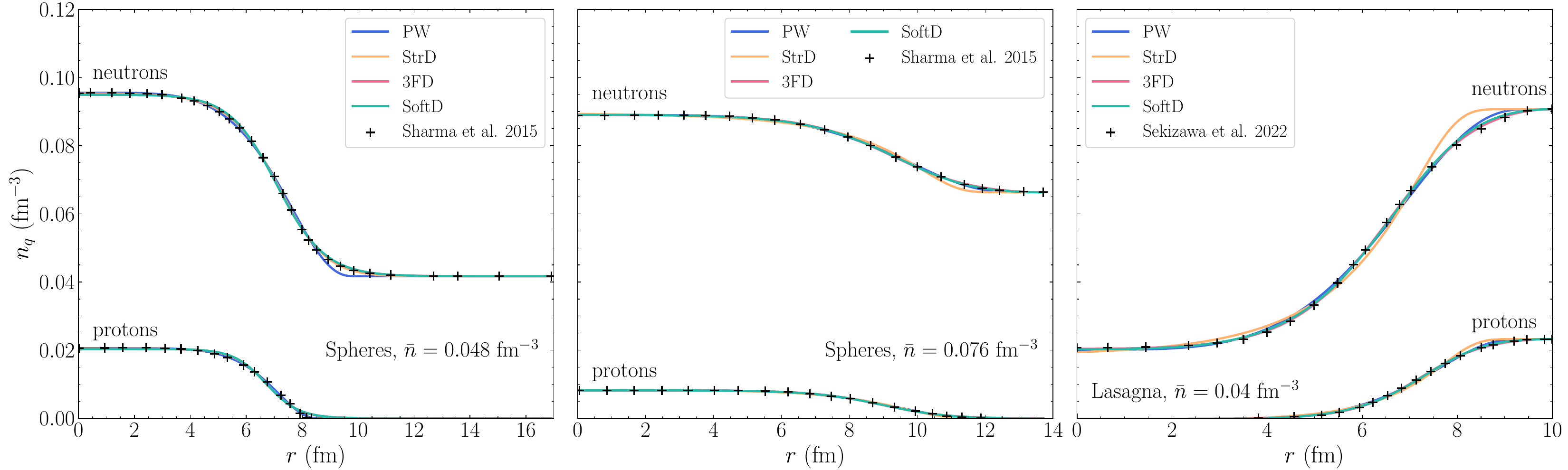}
	\caption{(left panel) Nucleon density distributions for spherical clusters at mean baryon density $\bar{n}=0.0475$ fm$^{-3}$. The results of Ref.~\cite{Sharma_Centelles+15} are represented by crosses, while the curves correspond to fits obtained with different parametrizations: PW (blue), StrD (orange), 3FD (pink) and SoftD (green). (middle panel) The same but for $\bar{n}=0.076$\,fm$^{-3}$. (right panel) The nucleon density distributions for the lasagna phase at $\bar{n}=0.04$ fm$^{-3}$ with $Y_\mathrm{p}=0.1$. Crosses are results from Ref.~\cite{Sekizawa+22}.}
	\label{fig:flex}
\end{figure}

In practice, we have fixed  for each parametrization the background density and the radius of the clusters
from the data and freely adjusted the remaining parameters $C_q, a_q$, and $n_{\mathrm{\Lambda}q}$. 
The resulting fits are presented in Fig.~\ref{fig:flex}. The two new parametrizations (3FD and SoftD) are able to accurately 
reproduce the equilibrium nucleon density distributions of Refs.~\cite{Sharma_Centelles+15} (although we show only the results 
for spherical clusters, the fits are equally good for other nuclear shapes) and \cite{Sekizawa+22}. 
The agreement is less satisfactory for the parametrization~\eqref{eq:PW}, for which the 
variations of the neutron densities appear too sharp, especially at low densities 
$\bar{n}$ (see also Ref.~\cite{Arponen72}). Although the StrD parametrization reproduces equally well the profiles
for spherical clusters at low densities, the quality of the fit deteriorates with increasing density. This 
parametrization fails to properly describe the diffuse nuclear surface and the nuclear distributions near
the border of the WS cell: imposing the vanishing of all the derivatives at $\xi=R$ appears too restrictive. 

These comparisons show that the two new parametrizations SoftD and 3FD are very well suited for describing nuclear pasta. 
The good fits to the self-consistent HF calculations of Ref.~\cite{Sekizawa+22} suggest that a fairly accurate description of lasagna could be already achieved at the ETF level.

\section{Vanishing of spaghetti}\label{appC}

\begin{figure}
	\includegraphics[width=\columnwidth]{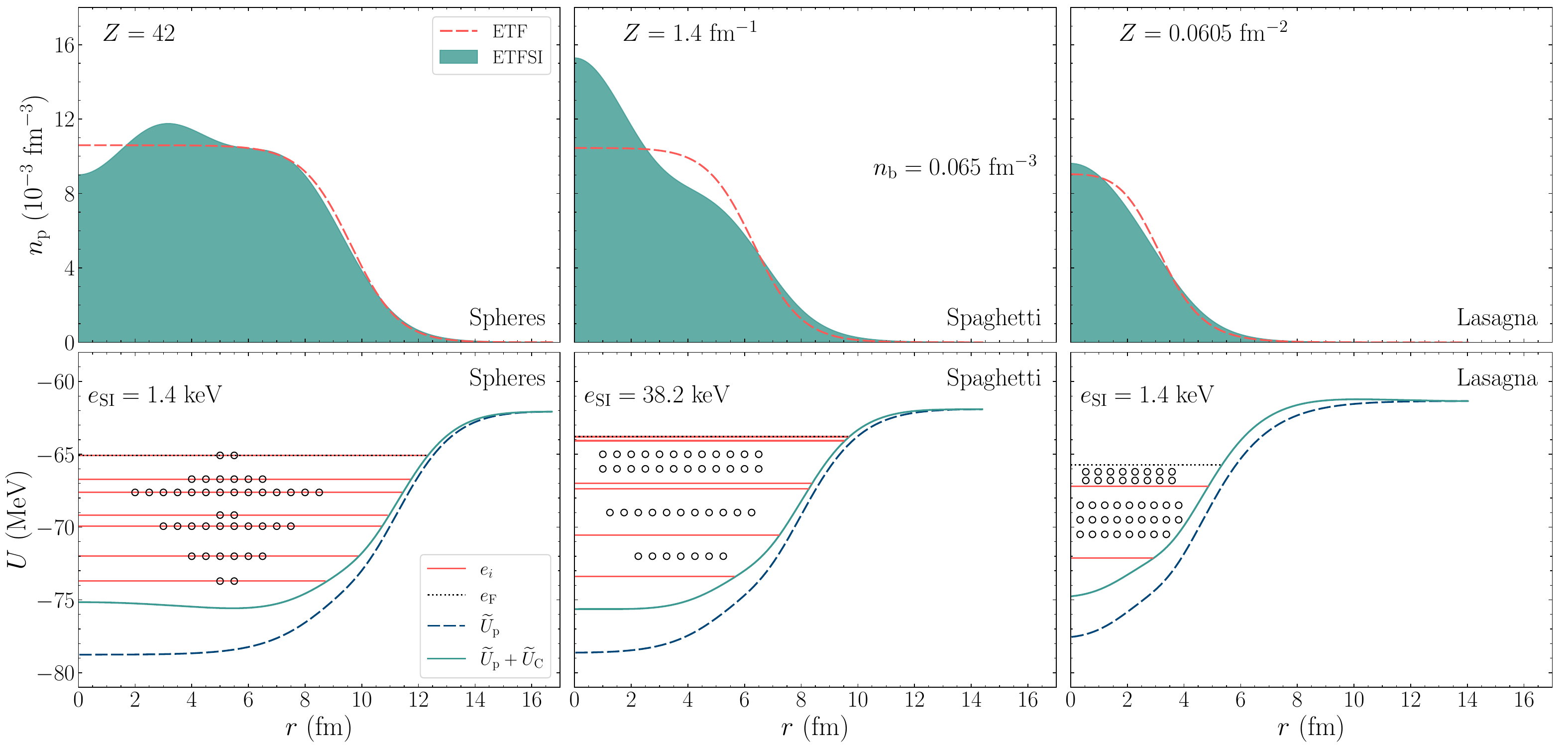}
	\caption{(top panels) Proton density distributions, as obtained within the ETFSI (filled area) or ETF (dashed lines) method,  for spheres, spaghetti, and lasagna at $\bar{n}=0.065\fm$. (bottom panels) The central (dashed navy lines) and total (solid green) proton mean-field potential are constructed from the ETF profiles. The energy levels are shown by red lines and the Fermi energy by black dotted lines. For spheres, the appropriate number of particles on each level is placed. For spaghetti and lasagna, as energy between the levels is filled by continuum states, the `nominal' number of particles between levels is shown. See text for details. }\label{fig:pot}
\end{figure}

The striking conclusion of Ref.~\cite{Pearson+22} was the complete vanishing of spaghetti 
within the ETFSI approach, a result confirmed here using new parametrizations of the nuclear 
profiles. To better understand this rather unexpected feature, we analyze more closely the 
role of the SI correction, focusing on the density $\bar{n}=0.065\fm$, for 
which spaghetti dominates over spheres and lasagna at the ETF level. We consider
here only the SoftD 
parametrization. For each pasta shape, we now fix $Z$ to the optimum value obtained within the 
ETF approach. The proton density distributions obtained at the ETF and ETFSI levels are plotted 
in Fig.~\ref{fig:pot}. For spheres and lasagna, results are essentially the same as in Fig.~\ref{fig:prof65} 
since the optimum $Z$ values are hardly altered by the SI corrections. For spaghetti, however, the 
proton density distribution obtained at the ETFSI level is significantly different than the one 
shown in Fig.~\ref{fig:prof65} due to the drop in $Z$ there. In particular, the distribution exhibits 
a bump resembling the one found in the HF+BCS calculations of Ref.~\cite{Gogelein&Muther07}. Such 
kind of profiles reflect the peculiarities of the single-particle wave functions of occupied 
proton states, which in turn are determined by the shape of the mean-field potentials. 
As shown 
in the bottom panels of Fig.~\ref{fig:pot}, the potential for spaghetti is as deep as for spheres
but significantly narrower although wider than for lasagna. This can be easily understood if we 
imagine the transformation of a spherical cluster into spaghetti or lasagna by stretching it and keeping 
the cell volume fixed and without changing the number of neutrons and protons (for a given mean 
baryon number density $\bar{n}$, the proton fraction is essentially determined by bulk properties, 
namely the symmetry energy, independently of the nuclear shape). 

The occupied proton states are illustrated in Fig.~\ref{fig:pot}. For spherical clusters, the 
energy of each level is discrete and can accommodate $2(2\ell+1)$ particles, where $\ell$ is 
the orbital quantum number (the factor of two accounts for the two possible spin projections 
recalling that we neglect the spin-orbit coupling). The situation is different for spaghetti and lasagna. The 
energy levels characterized by the set of quantum numbers $\nu$ can be written as~\cite{Pearson+22} 
\beqy\label{eq:levels}
\epsilon_\nu=&e_\mu+\dfrac{\hbar^2 k_z^2}{2m_p},& \text{ for spaghetti}\\
\epsilon_\nu=&e_\mu+\dfrac{\hbar^2 k_\perp^2}{2m_p},& \text{ for lasagna} \quad,\label{eq:levels2}
\eeqy
where $k_z=\pi n/L$ and $k_\perp=\pi\sqrt{n^2+m^2}/L$ are the wave numbers associated with  
motion along the spaghetti and in the plane of lasagna respectively ($n$ and $m$ are integers and 
$L$ is the size of the pasta). In both cases, $e_\mu$ can only take discrete values ($\mu$ represents 
the principal quantum number). But the energies in between are now also allowed so that the Fermi energy $\epsilon_\mm{F}$ 
lies above the last occupied state $\mu$. More particles can fill in the interlevel spacing for 
lasagna than for spaghetti due to the degeneracy with the direction of motion (see also~\eqref{eq:levels}-\eqref{eq:levels2}). This situation is illustrated in Fig.~\ref{fig:pot}. Although 
the total number of protons in each phase is the same and is approximately $42$ (the length of 
spaghetti and the area of lasagna was set so as to obtain the same volume as for spheres), 
they populate only two discrete levels for lasagna, six for spaghetti and seven for spheres. 
On the other hand, the narrower potential for spaghetti and lasagna tends to shift the 
energy levels to higher values than for spheres. The relative importance of the SI correction 
arises from the competition between these two effects. Spaghetti is found to have the largest 
number of protons on the most loosely bound levels resulting in a substantially larger SI 
correction of $38.2$ keV per nucleon compared to 1.4 keV per nucleon for spheres and lasagna (note that in all 
cases these corrections still remain very small compared to the total ETF energy per nucleon
 - less than $0.5\%$). 

In other words, the large SI correction for spaghetti stems from the quantum mechanical necessity 
of fitting the bound protons inside the potential, whose shape is constrained by the minimization
of the ETF energy at given mean baryon density $\bar{n}$. This analysis also suggests that the 
ETF approach could lead to spurious inhomogeneous configurations, which would vanish in a 
fully quantum mechanical treatment if the associated potential is too narrow or too shallow 
to contain bound energy levels~\cite{Brau2004}.

\bibliography{literature_profiles}

\end{document}